\documentclass[a4paper]{aipproc}
\usepackage{psfig}

\def\lta{ \lower .75ex\hbox{$\sim$} \llap{\raise .27ex \hbox{$<$}} }

\layoutstyle{6x9}

\SetInternalRegister\hbadness{8000} 

\begin{document}

\title {High Energy Phenomena in Blazars}


\author{L. Maraschi}{
  address={Osservatorio  Astronomico di  Brera, via Brera 28, 20121 Milano, 
Italy},
  email={maraschi@brera.mi.astro.it},
  thanks={This work was commissioned by the AIP}
}



\copyrightyear  {2001}

\begin{abstract}
Advances in the capabilities of X-ray, gamma-ray and TeV telescopes have
brought new information on the physics of relativistic jets, which are
responsible for the blazar "phenomenon". In particular the broad band
sensitivity of the $Beppo$SAX satellite, extending up to 100 KeV has
allowed unprecedented studies of their hard X-ray spectra.  I summarize
here some basic results and present a unified view of the blazar
population, whereby all sources contain essentially similar jets despite
diversities in other properties, like the presence or absence of emission
lines in their optical spectra.  Blazars with emission lines are of
particular interest in that it is possible to estimate both the
luminosity of the jet and the luminosity of the accretion
disk. Implications for the origin of the power carried by relativistic
jets, possibly involving rapidly spinning supermassive black holes are
discussed. We suggest that emission line blazars are accreting at near
critical rates, while BL lacs, where emission lines are weak or absent
are highly subcritical.
\end{abstract}

\date{\today}

\maketitle

\section{Introduction}

The original argument which led to hypothesize supermassive black holes
as basic engines for the AGN phenomenon (Zeldovich \& Novikov 1964;
Salpeter 1964) was extremely "simple". A source of high luminosity which
varies rapidly must be very "compact" : since its size $R$ must be less
than $ct_{\rm var}$ the photon density $L / (4\pi R^2 c)$ is extremely
high. A source of high compactness (defined in adimensional terms as
$l=L\sigma _T/R mc^3$) must be very efficient as shown most clearly by
Fabian (1979) who derived the well known limit $\Delta L/\Delta t \lta
\eta ~ 10^{43} erg/sec^2$, where $\eta $ is the radiative
efficiency. Accretion onto black holes can be orders of magnitude more
efficient than ordinary nuclear reactions powering stars, with $\eta $ up
to 42\% (e.g., Rees 1984) but still limited to values < 1.

Thus there is a maximum compactness that any source powered by accretion
cannot exceed. However observationally some AGN violate even this
limit. Early results concerned the excessive brightness temperatures
inferred from the variability of compact radio sources.  More recently,
the observation of gamma-rays varying rapidly, from the same class of
sources exhibiting the fast radio variability, provided new independent
evidence of violation of the fundamental compactness limit (Maraschi et
al. 1993).  For this relatively small fraction of AGN, called blazars,
relativistic motion of the emitting plasma has been invoked in order to
reconcile the observed properties with basic physics (Blandford and
Rees,1978).

It is now generally accepted that the blazar ``phenomenon'' (highly
polarised and rapidly variable radio/optical continuum) is due to a
relativistic jet pointing close to the line of sight. 
An additional step towards unification is to  propose
that jets are basically similar in all  blazars ,
 despite diversities in other properties , most notably
the presence or absence of emission lines  in their optical spectra (flat
spectrum quasars (FSQ) vs. BL Lacs).  This hypothesis was put forward
by Maraschi \& Rovetti (1994) on the basis of the "continuity" of the
radio and X-ray luminosity functions of the two classes of blazars.

Here I will follow the point of view  that Quasars with Flat Radio
Spectrum (FSQs, which include OVVs and HPQs) and BL Lac objects belong to
a single population, in the sense that the nature of the central engine
is similar apart from differences in scaling. The plan is to start from a 
physical
comprehension of the phenomenology common to all blazars, in particular
of the broad band spectral energy distributions (SEDs) from radio to
$\gamma$-rays, with the goal of understanding eventually the role of more
fundamental parameters, like the central black hole mass, angular
momentum and accretion rate, in determining the properties of the jets
and of the associated accretion disks.

\section{The unified framework for the SEDs of blazars.}

It was noted early on that the SEDs of blazars exhibit remarkable
systematic properties (Landau et al. 1986, Sambruna et al. 1996).  The
subsequent discovery by the Compton Gamma Ray Observatory of gamma-ray
emission from blazars (a summary can be found in Mukherjee et al. 1997)
was a major step forward, showing that in many cases the bulk of the
luminosity was emitted in this band and questioning the importance of
previous studies of the SEDs at lower frequencies.

A systematic investigation on the SEDs of the main complete samples of
blazars (X-ray selected, radio-selected and Quasar-like, Fossati et
al. 1998) including gamma-ray data showed that the systematic trends
found previously indeed persisted, suggesting a continuity of spectral
properties (spectral sequence). A suggestive plot where sources from
different complete samples have been grouped in radio-luminosity decades
 (see Fossati et al. 1998 for a full descripton) is shown in Fig. 1.

\begin{figure} 
\centerline{\psfig{file=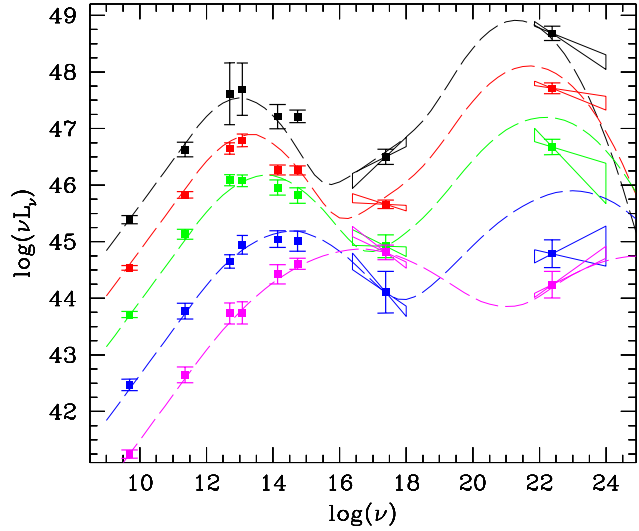,width=9.5cm} }
\vspace*{-1.0cm}
\caption{Average SEDs of the "merged" blazar sample
 binned according to radio
luminosity, irrespective of the original classification.  Empty asymmetric
 triangles represent uncertainties in spectral shapes
as measured in the X-ray and $\gamma$-ray bands. The overlayed dashed 
curves are analytic approximations obtained assuming that the ratio of the
two peak frequencies is constant and that the luminosity of the second peak is 
proportional to the radio luminosity (from Fossati et al.1998).}
\end{figure}

All the SEDs show two broad components with peaks in the $10^{13}
-10^{18}$Hz and $10^{21} -10^{25}$Hz ranges respectively. Both peaks
appear to shift to higher frequencies with decreasing luminosity. We will
call red and blue the objects respectively at the low and high frequency
extremes of the sequence .

Beamed synchrotron and inverse Compton emission from a single population
of relativistic electrons accounts well for the first and second peak
respectively in the observed SEDs.  Note that the relativistic particle
spectrum must be "curved" in order to explain the peaks observed in the
SEDs. The curvature is often modelled with a broken power law. The change
in spectral index must be quite large to explain the emission peaks and
the energy, $\gamma_b$, at which the change (or break) occurs 
corresponds to the energy
of electrons which radiate at the peak.

Homogenous models fail however to reproduce the SEDs in the radio to mm
range, where selfabsorption cuts off the contribution of the electron
population accounting for the higher energy emission (see also Kubo et
al. 1998).  In fact it is well known that at low frequency the observed
spectra are due to the superposition of different components located 
further down the jet, with selfabsorption turnover at lower
and lower  frequencies
(e.g., Konigl 1989).  These "external" regions, with scales of the order of parsecs
resolved by VLBI observations, are not considered in the present discussion
which refers to scales of the order of light days or smaller (in the observer's
frame). 

The homogenous model predicts that the synchrotron and IC emissions
should vary in a correlated fashion, since they derive from the same
electron population. In particular  radiation at frequencies near the two
peaks derives form electrons in the same energy interval (in the absence
of Klein-Nishina effects) therefore variations in these two
"corresponding" frequency ranges should be strongly correlated. 
 This has been verified at least in some well
studied objects (see below). 

Ghisellini et al. 1998 derived the physical parameters of jets of
different luminosities along the spectral sequence shown in Fig. 1
applying the above model and including seed photons of internal (SSC) as
well as external (EC) origin for the inverse Compton process.  The
results suggest that i) the importance of external seed photons increases
with increasing jet luminosity ii) the ``critical'' energy of the
radiating electrons decreases with increasing (total) radiation energy
density as seen in the jet frame.  The latter dependence is physically
plausible since the radiation energy density determines the energy losses
of relativistic particles and may limit the energy attained by particles
in shock acceleration processes.

In a broader perspective, if FSQs and BL Lacs contain "similar" jets (at
least close to the nucleus) as suggested by the continuity of the SEDs,
we still need to understand the differences in emission line properties.
Also in this respect continuity could hold, in the sense that the
accretion rate may decrease continuously along the sequence but the
emission properties of the disk may not simply scale with the accretion
rate.

\section{Studies of individual objects.}

It is impossible to review here all the important multifrequency studies
of selected objects often triggered by flares (e.g. Bloom et al. 1996,
 Kubo et al. 1998, Tagliaferri et al. 2000, 2001, Sambruna 2000 and
references therein, ) We discuss below only three cases which are among those with
the best data collection and can be considered representative of the behaviour 
of sources at the red and blue ends of the blazar sequence. 

\subsection {3C 279}

This source is a prototype of red blazars and the first one to be
discovered as a powerful gamma-ray emitter.  Its spectral energy
distribution illustrates well the presence of two main continuum
components, the first one peaking in the IR, the second one in the
gamma-ray range, attributed to the Synchrotron and inverse Compton
mechanisms respectively. Two SEDs with simultaneous optical, X-ray and
gamma-ray data obtained respectively in 1996 and 1997 are shown in
Fig. 2.  The 1997 X-ray data derive from observations with $Beppo$SAX
(Hartman et al. 2001, Maraschi et al. in preparation). The two SEDs
differ largely in brightness: that of 1996 is close to a historical
maximum, while that of 1997 is rather faint. Clearly the two components
of the SED vary in a correlated fashion (this is confirmed by a handful
of other observations with comparable simultaneous multifrequency
coverage) and the variability amplitude in the IR-optical branch is much
less than at gamma-rays.  This was predicted by the SSC model (Ghisellini
\& Maraschi 1996) but the physical parameters derived using a homogeneous
SSC model are inconsistent with the rapid gamma-ray variations observed.
Applying the EC model for gamma-ray production yields acceptable
parameters. However the photon field surrounding the jet is not expected
to vary rapidly except under special conditions (Ghisellini \& Madau
1996, Bednarek 1998, B\"{o}ttcher \& Dermer 1998,). The large amplitude
variability in gamma-rays can then only be attributed to a variation in
the bulk flow velocity as illustrated in Fig 2.

\begin{figure} 
\centerline{\psfig{file=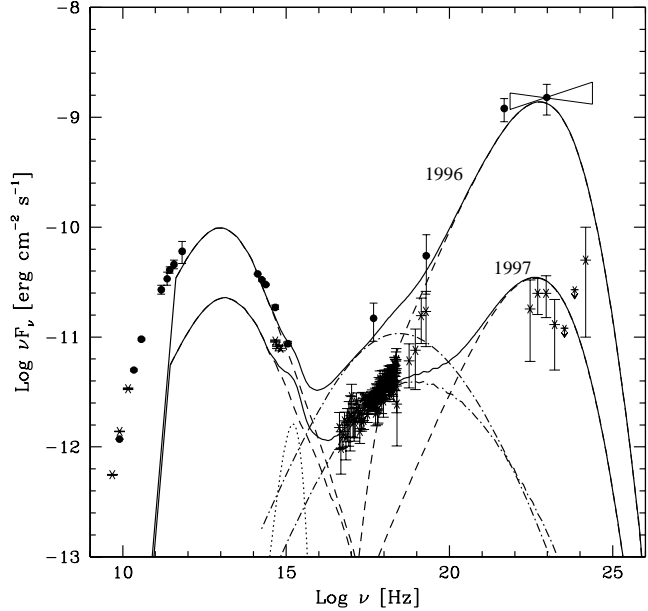,width=9 cm}}
\caption{Quasi-simultaneous SEDs of the quasar 3C279 obtained at two
epochs 1996 and 1997. The 1996 SED represents a historical maximum.  The
X-ray data of 1997 were obtained with {\it Beppo}SAX.  The continuous
lines represent synchrotron plus Inverse Compton models computed to
reproduce the observations at the two epochs (see text). For both epochs
subcomponents of the Inverse Compton emission are shown as dot-dashed
(SSC) and dashed (EC) lines respectively. The dotted component is a
Blackbody approximating the estimated emission from an accretion disk,
assumed to be constant.  The models for the two epochs differ mainly in
the value of $\Gamma_{bulk}$}
\end{figure}

The two spectral states have been reproduced using the same theoretical
model (e.g. Tavecchio et al. 2000) varying only the bulk Lorentz factor
$\Gamma$ of the emitting plasma (Maraschi et al. 2001 in prep). The
contributions of synchrotron photons and of external photons to the
Inverse Compton emission are shown separately (dot dashed and dashed
lines respectively. While this picture is probably still oversimplified,
it fits nicely with a recently proposed scenario derived from the
"internal shock" model developed for Gamma-Ray Bursts. In the latter
(Spada et al. 2001, in press), most of the variability is attributed to
the collisions of plasma sheaths moving along the jet with different
$\Gamma$s.  This scenario is very promising for explaining the full range
of variability of 3C 279.

For red blazars the study of the synchrotron component is difficult,
because the peak falls in the poorly covered IR - FIR
range. Furthermore the study of the gamma-ray component in the MeV-GeV
region of the spectrum has been difficult in the last few years due to
the loss of efficiency of EGRET and is now impossible after reentry of
CGRO. Substantial progress will have to await the launch of new gamma-ray
satellites, like AGILE planned by the Italian Space Agency (ASI) and
GLAST by NASA.

\subsection {Mkn 501, Mkn 421}

In the last years high energy observations have concentrated on blue
blazars.  For several sources of this class the Synchrotron component
peaks in the X-ray band, where numerous satellites can provide good
data. In few bright extreme BL Lac objects the high energy $\gamma$-ray
component is observable from ground with TeV telescopes (for a general
account see Catanese \& Weekes 1999). In these particular cases the
contemporaneous X-ray/TeV monotoring demonstrated well the correlation
between the Synchrotron and the IC components. Dramatic TeV flares
exhibited by Mkn 501 were accompanied by exceptional outbursts
in X-rays where {\it in the brightest state the peak of the synchrotron 
spectrum reached 100 KeV} as  observed by {\it Beppo}SAX (Pian et al. 1998).
Similar behaviour was observed with RXTE (Sambruna et al. 2000,
 Catanese \& Sambruna 2000).

Another important case is that of Mkn 421 for which a rapid flare 
(timescale of hours) was
observed simultaneously in the TeV and X-ray bands by the Whipple
observatory and by the {\it Beppo}SAX satellite in 1998. This observation
probed for the first time the existence of correlation on short 
time scales (Maraschi et al. 1999).  An associated intensive
multiwavelength campaign (involving EUVE ASCA RXTE and the CAT, HEGRA and
Whipple observatories) organized by Takahashi showed that TeV variations
correlate with X-rays also on longer timescales and confirmed the trend
of spectral hardening with increasing X-ray intensity (Takahashi et
al. 1999).

\begin{figure}
\hspace{0.2truecm}
\psfig{file=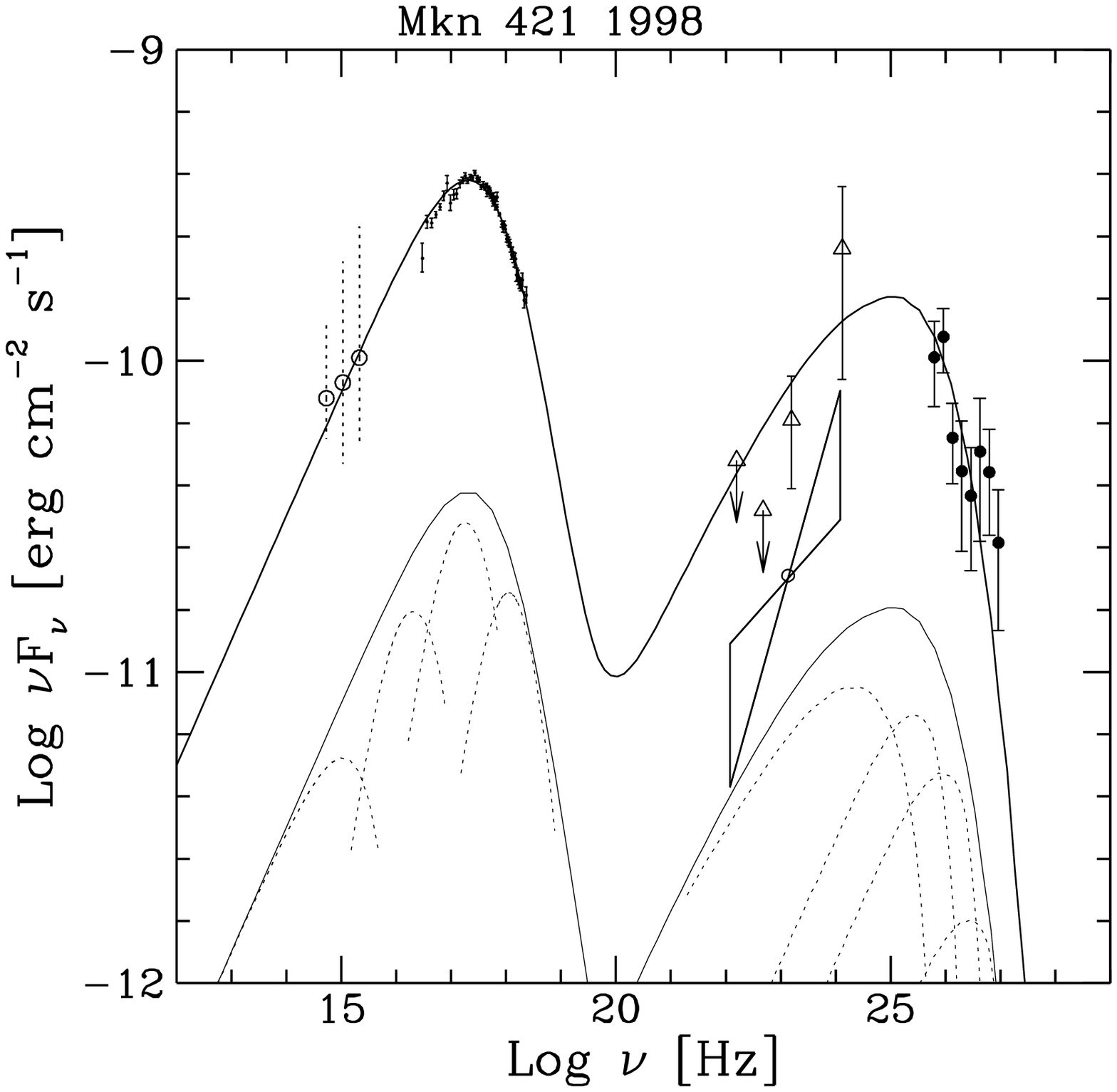,width=6.5 cm}
\vspace{-6.5truecm}
\hspace{.1truecm}
\psfig{file=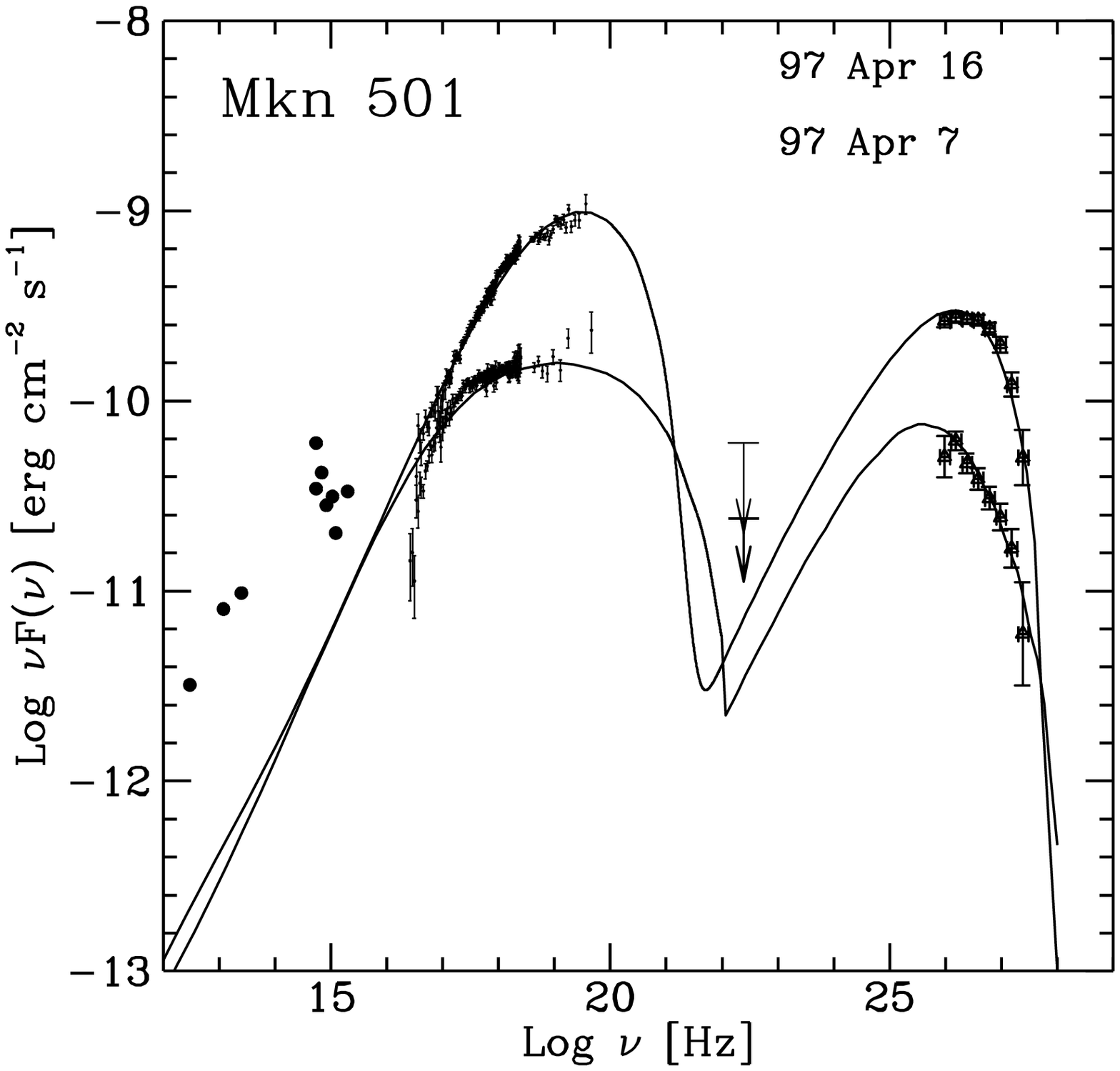,width=6.5 cm}
\caption{{\it Left:} Overall SED of Mkn 421 obtained from
observations taken in 1998 April (from Maraschi et al. 1999). The X-ray and TeV
data are exactly simultaneous. The solid line is the spectrum computed with the
SSC model. Below the actual model, subcomponents due to electrons in 4 fixed 
energy intervals are shown
both for the synchrotron and SC processes.
 {\it Right:}
Overall SED of Mkn 501 observed simultaneously by {\it Beppo}SAX and CAT
during the major flare in April 1997 (From Tavecchio et al. 2000, in
press). The solid lines are the spectra computed with the
SSC model. The models for the high and low state differ only in the
value of $\gamma_b mc^2$, the energy of the electrons radiating at the
peak of the synchrotron component. 
For both sources the observational  constraints on the two peaks 
allow to obtain robust estimates of the physical parameters of the jet.}
\end{figure}

When the position of the synchrotron and SSC peaks can be well determined
observationally, as is possible in this type of sources, robust estimates
of the physical parameters of the jet can be obtained (e.g. Tavecchio et
al. 1998). This was done for both Mkn 421 and Mkn 501 as illustrated in
Fig. 3 (Maraschi et al. 1999, Tavecchio et al., in press). 
For Mkn 421 subcomponents due to electrons in 4 fixed energy intervals
are shown in Fig. 3  both for the synchrotron and SSC process, in order to
give an intuitive view of the correlation between different energy
ranges. Note for Mkn
501 the definite change in the TeV spectra measured by the CAT group
(Djannati-Ata{\i} et al. 1999) indicating a shift of the IC peak
consistent with the change of the X-ray spectrum.

As a result of these model fits to different states with the
simultaneous constraint on the X-ray and TeV spectra, we can confidently
deduce that the flares are due to an increase of the critical electron
energy $\gamma_b$ rather than to a variation of the bulk Lorentz factor
 $\Gamma$ as suggested for 3C279. These "modes"
of variability may represent a significant difference between red and
blue blazars.

\section{Jet power vs. accretion power}

We now turn to discuss luminous blazars with
emission lines. These fall at the
high-luminosity end of the sequence, with the Synchrotron peak in the FIR
region. In these sources the beamed X-ray emission is believed to be produced
through  IC scattering between soft photons external to the jet
(produced and/or scattered by the Broad Line Region) and {\it relativistic 
electrons
at the low energy end of their energy distribution}. 
It is important to stress that the broad band sensitivity of {\it Beppo}SAX
allowed to measure the X-ray spectra from 0.3 up to 100 KeV for a number of 
these objects. Note that for this type of sources the 
hard X-ray emission has luminosity comparable to that measured in gamma-rays,
due to the fact that the EC peak falls in between the two ranges. 

Measuring the X-ray spectra and adapting a broad band model to their SEDs
yields reliable estimates of the total number of relativistic particles
involved, which is dominated by those at the lowest energies.  This is
interesting in view of a determination of the total energy flux along the
jet (e.g. Celotti et al. 1997, Sikora et al. 1997). The  total " kinetic"
power of the jet can be written as:
\begin{equation}
P_{\rm jet}=\pi R^2 \beta c \,U \Gamma ^2
\end{equation}
\noindent
where $R$ is the jet radius, $\Gamma$ is the bulk Lorentz factor and $U$
is the total energy density in the jet, including radiation, magnetic
field, relativistic particles and eventually protons. If one assumes that
there is 1 (cold) proton per relativistic electron, the proton
contribution is usually dominant.

In high luminosity blazars the UV bump is often directly observed and/or
can be estimated from the measurable emission lines, yielding direct
information on the accretion process in the hypothesis that the UV
emission derives from an accretion disk. {\it Thus the relation between
accretion power and jet power can be explored}. 
This approach was started
by Celotti et al. (1997) but their estimates of $P_{\rm jet}$ were
obtained applying the SSC theory to VLBI radio data which refer to pc
scales, much larger than the region responsible for the high energy
emission ($10^{-2} - 10^{-3}$ pc).

We took advantage of  {\it Beppo}SAX data for a number of emission line blazars
deriving their jet powers as described above. The SEDs for three objects
together with the models computed to represent the data are shown in Fig 4
(from Tavecchio et al. 2000).
We  have preliminary results for 6 other sources with similar characteristics,
all observed with  {\it Beppo}SAX.
We further consider blazars with less prominent emission lines,
for which we had previuos good quality {\it Beppo}SAX and multifrequency data,
namely 3C 279, BL Lac, ON231 (Tagliaferri et al. 2001, Tagliaferri et al. 2000)
plus Mkn 501 and Mkn 421 discussed above.
 
\begin{figure}
\centerline{\psfig{file=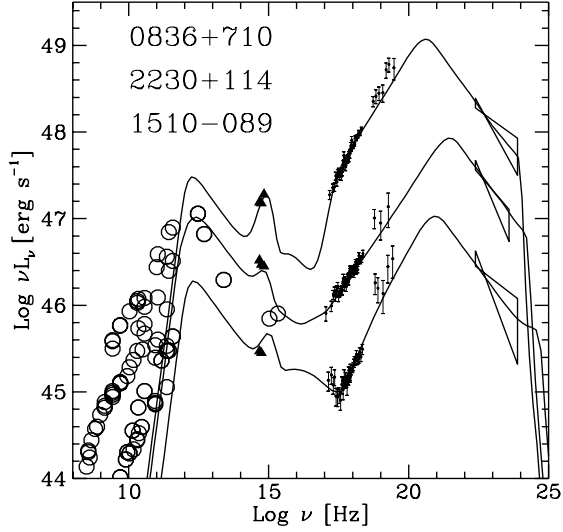,height=9.cm}}
\vspace{-0.8truecm}
\caption{Overall SEDs of three powerful emission-lines Blazars (from
Tavecchio et al. 2000). The continuous lines are theoretical models
computed to account for the jet emission, plus a blackbody component.
  The objects are characterized by the presence of
a strong UV-bump, allowing the determination of the luminosity of the
accretion disk as well as the luminosity and power of the jet from the
model fits to the non thermal emission.}
\end{figure}

In all cases we estimated physical parameters by means of a homogeneous
SSC+EC model and derived accordingly the kinetic power of the jet
including 1 cold proton per electron, $P_{\rm jet}$, as well as the total
luminosity radiated by the jet in the observer frame ($L_{\rm jet}$).
The luminosity of the disk could be estimated for all objects except the
latter three BL Lac, for which we could set only upper limits on the
luminosity of their putative accretion disks. For 3C 279 and BL Lac, the
presence of broad Ly$_{\alpha}$ and H$_{\alpha}$ respectively allowed to
estimate the ionizing continuum (e.g. Corbett et al. 2000).

In Fig 5a the derived radiative luminosity $L_{\rm jet}$ and kinetic
power of the jet $P{\rm jet}$ are compared. The ratio between these
two quantities gives directly the "radiative efficiency" of the jet,
which turns out to be $\eta\simeq 0.1$, though with large scatter.  The
line traces the result of a least-squares fit: we found a slope $\sim 1.3$,
indicating a decrease of the radiative efficiency with decreasing
power.

In Fig. 3b we  compare  the luminosity of the jet, $L_{\rm jet}$,
which is a {\it lower limit} to $P_{\rm jet}$, with the luminosity of 
the disk, $L_{\rm disk}$.

A first important result is that on average the minimal power
transported by the jet is {\it of the same order} as the luminosity
released in the accretion disk. This result poses an important constraint
for models elaborated to explain the formation of jets.

Two main classes of models consider either extraction of rotational
energy from the black hole itself or magnetohydrodynamic winds associated
with the inner regions of accretion disks. Let us parametrize the two
possibilities as follows.  Blandford \& Znajek (1977) summarize the
result of their complex analysis of extraction of rotational energy from
a black hole in the well known expression: 
\begin{equation} P_{BZ}\simeq
B_0^2 r_g^2 a^2 c \,\,\,\,\, \end{equation} 
Assuming maximal rotation for
the black hole ($a=1$), the critical problem is the estimate of the
intensity reached by the magnetic field threading the event horizon,
which must be provided by the accreting matter. 
 Using a spherical free
fall approximation with $B_0^2/ 8\pi \simeq \rho c^2$ one can write:
\begin{equation} P_{BZ}\simeq g \dot{M}c^2 \,\,\,\,\, \end{equation}
where $P_{acc}=\dot{M}c^2$ is the accretion power and $g$ is of order 1
in the spherical case, but in fact it is a highly uncertain number since
it also depends on the field configuration. 

 Several authors have
recently discussed this difficult issue in the case of an accretion disk:
the arguments discussed by Ghosh \& Abramovicz (1997) (GA; see also
Livio, Ogilvie \& Pringle 1999) plus equipartition within an accretion
disk described by the Shakura and Sunyaev (1973) model lead to $g \simeq
1$ when gas pressure dominates. However at high accretion rates,
when radiation pressure dominates, the pressure and consequently the
estimated magnetic field do not increase further with $\dot{M}$ but
saturate at the transition value. The estimates of $P_{BZ}$ derived by GA
for various values of the mass of the central black hole are compared
with the values of $L_{jet}$ and $L_{disk}$ in Fig 5b. The accretion rate
which appears in the formulae of GA has been converted into a disk
luminosity using an efficiency $\epsilon \simeq 0.1$, while 100\%
radiative efficiency has been assumed for the jet.  Clearly the model
fails to explain the large power observed in the jets of bright quasars,
even for BH masses ($M\sim 10^9 M_{\odot}$).
Different hypotheses on the structure of the flow near the black hole,
for instance frame dragging by the rotating hole may however increase $g$ to
values even larger than 1 (Meier 1999, Krolik 1999; Acceleration and
collimation of cosmic jets (session 6), this volume). 

 As argued  by Livio et
 al. the accretion flow itself may power jets through a hydromagnetic
 wind. However for consistency only some fraction $f \dot{M}c^2$ can be
 used to power the jet.  Further recall that the luminosities observed
 from the jet and disk are related to their respective powers by
 efficiency factors $L_{jet}=\eta P_{jet}$ ; $ L_{disk}=\epsilon P_{acc}$.

Using the condition that $P_{jet} \leq (P_{BZ} + f P_{acc} )$
together with the previous relations we finally find
\begin{equation} 
L_{jet} \leq \frac{\eta (g+f)} {\epsilon }L_{disk}.
\end{equation}

The data we have used suggest $L_{jet} \simeq L_{disk}$ at high
luminosities and $L_{jet} > L_{disk}$ at intermediate and low
luminosities. 

At the high luminosity end the observed luminosities are extremely
large.  For a disk luminosity of $10^{47}$ erg s$^{-1}$ a mass of $10^9$
$M_{\odot}$ is implied if the disk is close to the Eddington luminosity,
which corresponds to an accretion rate of $10$ $M_{\odot}$ y$^{-1}$ for
$\epsilon = 10^{-1}$.  It seems then implausible that such disks could be
low efficiency radiators.  Assuming that $\eta = 10^{-1}$ as estimated
above (Fig 5a), the near equality of $L_{jet}$ and $L_{disk}$ requires
$g$ or $f$ or both to be of order 1.

On the other hand, a dominance of $L_{jet}$ over $L_{disk}$ at lower
luminosities could be attributed to a lower value of $\epsilon << 0.1$
which may be expected if the accretion rate is largely sub-Eddington
(e.g., Blandford 1990).  In the latter case the range in
luminosities spanned by Fig 5b should be mainly a range in accretion rates 
rather than a range in black hole masses. 
For instance the minimum jet powers of three of the BL
Lacs in Fig 5b are around $10^{44}$ erg/s which suggests $P_{jet}\simeq
10^{45}$ requiring a mass of $10^7$ for critical accretion rate. Since
the disk luminosity is less than $10^{42}$, if the accretion rate is 1\%
Eddington the implied mass is again $10^9$ $M_{\odot}$.  This scenario is attractive
(see also Cavaliere \& Malquori 1999) and could be verified
observationally if the mass of the central object can be determined
independently.  In fact for low luminosity objects
the  velocity dispersion close to the
core of the galaxy, indicative of the central black hole mass
 (e.g. Ferrarese et al. 2000) should be measurable.

\begin{figure}
\hspace{0.2truecm} 
\psfig{file=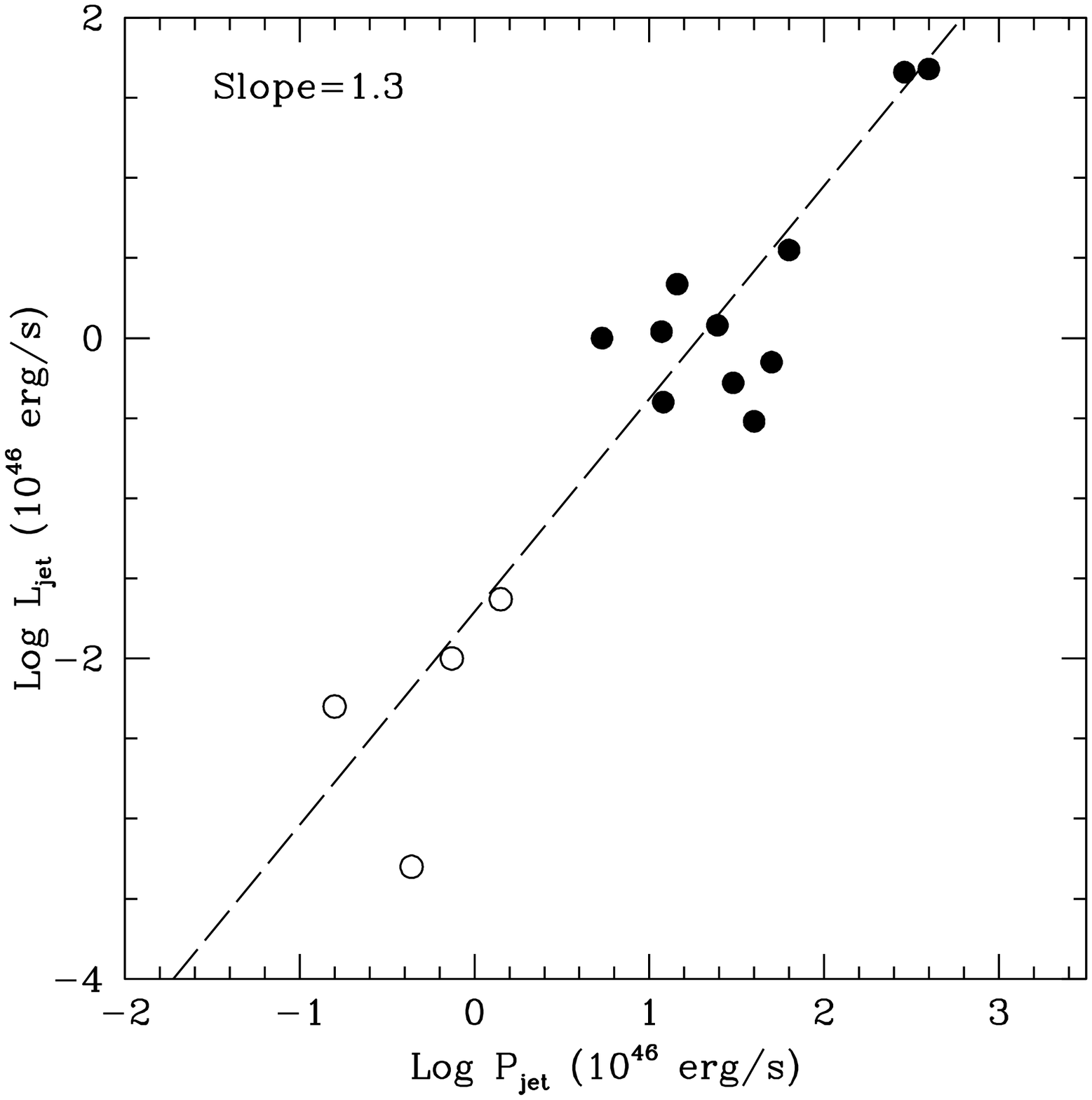,width=6.cm}
\vspace{-6.5truecm}
\hspace{.1truecm}
\psfig{file=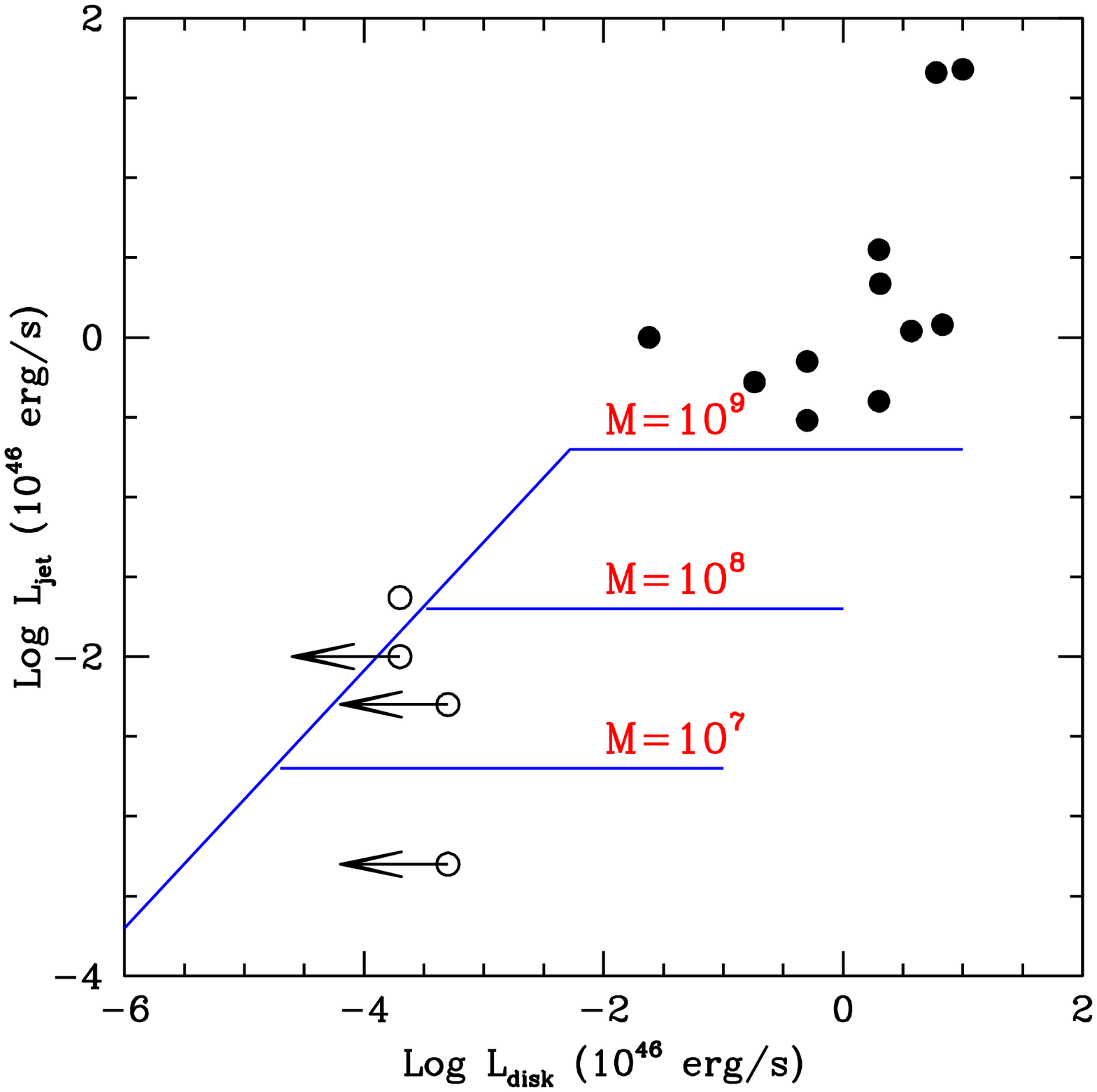,width=6.cm}
\caption{{\it Left:} Radiative luminosity vs. jet power for the sample of
Blazars discussed in the text (open circles represent BL Lac
objects). The dashed line indicates the least-squares fit to the
data. {\it Right:} Radiative luminosity of jets vs disk luminosity. The
solid lines represent the {\it maximum} jet power estimated for the
Blandford \& Znajek model for black holes with different masses (in 
Solar units).}
\end{figure}

\section{Conclusions}
The study of blazars yields unique information on the physical conditions
and emission processes in relativistic jets.  A unified approach is
possible whereby the jets in all blazars are similar and their power sets
the basic scale.  While the phenomenological framework is suggested to be
"simple" (e.g. "red" blazars are highly luminous, have low average
electron energies and emit GeV gamma-rays while "blue" blazars have low
luminosity, high average electron energies and emit TeV gamma-rays) we do
not yet know what determines the emission properties of jets of different
power nor what determines the jet power in a given AGN. We suggest that 
the basic parameter may be the accretion rate rather than the black hole mass
that is all blazars contain very massive black holes and the lower luminosity
ones are accreting at sub-Eddington rates. An observational verification
of this hypothesis may come from black hole mass determinations in the nearest,
lowest luminosity BL Lacs.

\vskip 0.4 cm

\bibliography{sample}

\noindent
Blandford, R.D., \& Znajek, R.L., 1977, MNRAS, 179, 433\\
Blandford, R.D., \& Rees, M.J., 1978, Pittsburgh Conf. on BL Lac
Objects, p. 341-347.\\
Blandford, R.D. 1990, Saas-Fee Advanced Course 20. Lecture Notes
1990. Swiss Society for Astrophysics and Astronomy, XII, 280 pp. 97.\\
Bloom, S. D. et al.1997, ApJ, 490, L145\\ 
Bednarek, W., 1998, A\&A, 386, 123\\
B\"{o}ttcher, M. \& Dermer, C.D., 1998, ApJ, 501, 51\\
Catanese, M., \& Weekes, T.C., 1999, PASP, 111, 1193\\
Catanese, M.\& Sambruna, R.\ M.\ 2000, ApJ, 534, L39\\ 
Cavaliere, A. \& Malquori, D. 1999, ApJ, 516, L9\\
Celotti, A., Padovani, P., \& Ghisellini, G. 1997, MNRAS, 286, 415\\ 
Corbett, E. A., et al. 2000, MNRAS, 311, 485\\ 
Djannati-Ata{\i} , A. et al. 1999, A\&A, 350, 17 \\
Fabian, A.C. 1979, Proc. Royal. Soc., 366, 449\\
Fossati, G., et al. 1998, MNRAS, 299, 433\\
Ghisellini, G., \& Madau, P., 1996, MNRAS, 280, 67\\
Ghisellini, G., \& Maraschi, L., 1996, ASP Conf. Series, 110, 436\\
Ghisellini, G., et al. 1998, MNRAS, 301, 451\\
Ghosh, P., \& Abramowicz, M.A. 1997, MNRAS, 292, 887\\ 
Krolik, J.H., 1999, ApJ, 515, L73\\
Hartman, R.C. et al. 2001, ApJ, in press\\
Konigl, A., 1989, in ``BL Lac Objects''..., 321\\ 
Kubo, H., et al. 1998, ApJ, 504, 693\\
Landau, R., et al. 1986, ApJ, 308, 78\\
Livio, M., Ogilvie, G. I. \& Pringle, J. E. 1999, ApJ, 512, 100\\
Maraschi, L. \& Rovetti, F. 1994, ApJ, 436, 79\\
Maraschi, L., et al. 1999, ApJ, 526, L81 \\
Meier, D. L. 1999, ApJ, 522, 753\\
Mukherjee, R., et al. 1997, ApJ, 490, 116\\ 
Pian, E., et al. 1998, ApJ, 492, L17\\
Rees, M.J., 1984, ARA\&A, 22, 471\\
Salpeter, E. E. 1964, ApJ, 140, 796\\ 
Sambruna, R.,M., Maraschi, L., \&Urry, C.M., 1996, ApJ, 463, 444\\
Sambruna, R. M., 2000, AIP, 2000. AIP Conference Proceedings, Vol. 515, 19\\
Sambruna, R. M., et al. 2000, ApJ, 538, 127\\ 
Shakura, N. I. \& Sunyaev, R. A. 1973, A\&A, 24, 337\\ 
Sikora, M., et al. 1997, ApJ, 484, 108\\  
Takahashi, T., et al. 1999, Astroparticle Physics, 11, 177\\ 
Tagliaferri, G. et al. 2000, A\&A, 354, 431\\
Tagliaferri, G.et al. 2001, A\&A, 368, 38 \\
Tagliaferri, G. et al. 2001, to appear  Proceedings of the
Conference ``X-ray Astronomy'99: Stellar Endpoints, AGN and Diffuse
Background", 2000, Astrophysical Letters and Communications, in press\\
Tavecchio, F., Maraschi, L., \& Ghisellini, G., 1998, ApJ, 509, 608\\ 
Tavecchio, F., et al. 2000, ApJ, 543, 535\\
Zeldovich Y.B., \& Novikov, I.D. 1964 Usp. Fiz. Nauk 84, 377\\

\end{document}